\documentclass[twocolumn,letterpaper,aps,prc,superscriptaddress,nofootinbib,showpacs,floatfix]{revtex4-1}% PRC format
\usepackage{graphicx}
\usepackage{comment}
\usepackage{setspace}
\usepackage{amsmath}
\usepackage{amssymb}
\usepackage[textwidth=16cm,textheight=22cm]{geometry}
\usepackage{color}
\usepackage{indentfirst}
\usepackage{xspace}
\usepackage{hyperref}
\usepackage{verbatim}
\usepackage{epstopdf}

\begin{document}
\title{Multiplicity distribution and normalized
  moments in p$-$p collisions at LHC in forward rapidity using Weibull model}

\author{Ranjit~Nayak}
\email{ranjit@phy.iitb.ac.in}
\author{Sadhana~Dash}
\email{sadhana@phy.iitb.ac.in}
\affiliation{Indian Institute of Technology Bombay, Mumbai,
  India-400076}

\date{\today}

\begin{abstract}
The measured charged particle multiplicity distribution in the forward
rapidity region of the p$-$p collisions at $\sqrt{s}$ = 0.9 TeV, 7 TeV
and 8 TeV at the LHC have been described using the Weibull distribution 
function. The higher order (up to $5^{th}$ order ) normalized moments
and the factorial moments are also calculated using the extracted
parameters. The multiplicity distributions in forward region are
observed to be well described by the Weibull regularity and
the higher order moment calculations confirm the violation of KNO
scaling that has been observed at mid-rapidity and lower energies. 
The Weibull parameters and moments for p$-$p collisions at 
$\sqrt{s}$ = 13 TeV are also estimated for the forward region. 
\end{abstract}

\maketitle
\section{Introduction}
%%%%%%%%%%%%%%%%%%%%%%%%%%%%%%%%%%%%%%%%%%%%%%%%%%%%%%%%%%%%%%%%%%%%%
Recently, the ALICE (A Large ion collider Experiment) experiment at
LHC (Large Hadron Collider) has measured the charged
particle multiplicity distribution in wider pseudo-rapidity ranges for
different event classes, INEL(inelastic events), INEL $>$ 0 (same as INEL
with the presence of at least one charged particle for $|\eta| < 1.0$ and
NSD (non-singly diffractive events) \cite{alicefor}. This particular
measurement allowed extending the previous measurements of other 
LHC experiments to a wider $\eta$ coverage enabling to reach higher
multiplicity and kinematic phase space together with previous
measurement by CMS\cite{cms}, ALICE\cite{alicemid}, and LHCb\cite{lhcb}.
The forward region is also relevant in view of being sensitive to
interesting low Bjorken-x dynamics and multi-particle interactions\cite{mpi}.

The charged particle multiplicity distribution in full phase space for 
p$-$p collisions are influenced by the global conservation laws
(namely energy-momentum, charge etc). However, the distributions in
restricted phase space are less influenced by the global constraints
and are sensitive to the dynamics of particle production and local
fluctuations\cite{kittel,wolf}. The predictions of several
QCD-inspired Monte Carlo event generators such as PYTHIA
\cite{pythia8}, PHOJET\cite{phojet} etc. were used to compare the
data. It was found that these models underestimated the
data\cite{alicefor,cms,alicemid,lhcb}. Therefore, it becomes relevant to study the
distributions in different event classes and varied phase space
intervals to allow constraining the existing Monte-Carlo model
parameters and understand the underlying dynamics of multi-particle 
production. 
The multiplicity distribution at LHC energies are observed to deviate
from Poissonian shape and thus signals toward the existence of
correlations\cite{sarkisyan,sarkisyan1}. These multi-particle
correlations can be studied through
the normalized moments and factorial moments of the distribution. 
The study of moments is also sensitive to the KNO (Koba, Nielsen and
Olesen) scaling i.e the energy independence of moments of various
orders would imply the observance of KNO scaling\cite{kno}. 
In this work, Weibull distribution\cite{weib1,weib2} has been used to describe the
multiplicity distribution of charged particles in pp collisions at
$\sqrt{s}$ = 0.9 TeV, 7 TeV and 8 TeV as measured by ALICE experiment
in wider $\eta$ intervals (forward region) for different event
classes. The study was extended to calculate the  higher moments to
confirm the violation of KNO scaling as reported in previous
measurements for central rapidity regions\cite{cms,alicemid,weibmom}.
\section{Moments of Weibull Distribution}
\label{formulation}
The probability density distribution of Weibull for a continuous random variable $x$ is given by-
\begin{eqnarray}
f(x;\lambda , k) = \frac{k}{\lambda} \left(\frac{x}{\lambda}\right)^{(k-1)}e^{-(\frac{x}{\lambda})^{k}}
\label{e1}
\end{eqnarray}
where $k$ is the shape parameter and $\lambda$ is the scale parameter of the distribution.

The n$^{th}$ raw moment is given by:
\begin{eqnarray}
m_n = \lambda^n \, \Gamma\left(1 + \frac{n}{k}\right)
\label{e2}
\end{eqnarray}
The mean of the distribution, i.e. $m_1$ is denoted by 
$\langle x \rangle$  and is given by, 
\begin{eqnarray}
\langle x \rangle = \lambda \, \Gamma\left(1 + \frac{1}{k}\right)
\label{e2}
\end{eqnarray}
The n$^{th}$ factorial moment of a variable $x$ is defined as:
\begin{eqnarray}
f_n &=& \Big\langle \frac{x!}{(x - n)!} \Big\rangle \nonumber \\
    &=& \langle x(x-1)(x-2)...(x-n+1) \rangle
\label{e3}
\end{eqnarray}
The $n^{th}$ normalized moment $C_n$ and normalized factorial moments $F_n$ are  defined as:
\begin{eqnarray}
C_n = m_n/m_1^n ; \,\,\,\,\,\,\, F_n = f_n/m_1^n
\label{e4}
\end{eqnarray}
The first few normalized moments and factorial moments can be found at \cite{weibmom}.
%%%%%%%%%%%%%%%%%%%%%%%%%%%%%%%%%%%%%%%%%%%%%%%%%%%%%%%%%%%%%%%%%%%%%
\section{Analysis and Results}
\begin{figure*}
\includegraphics[width=5.2in,height=5.8in]{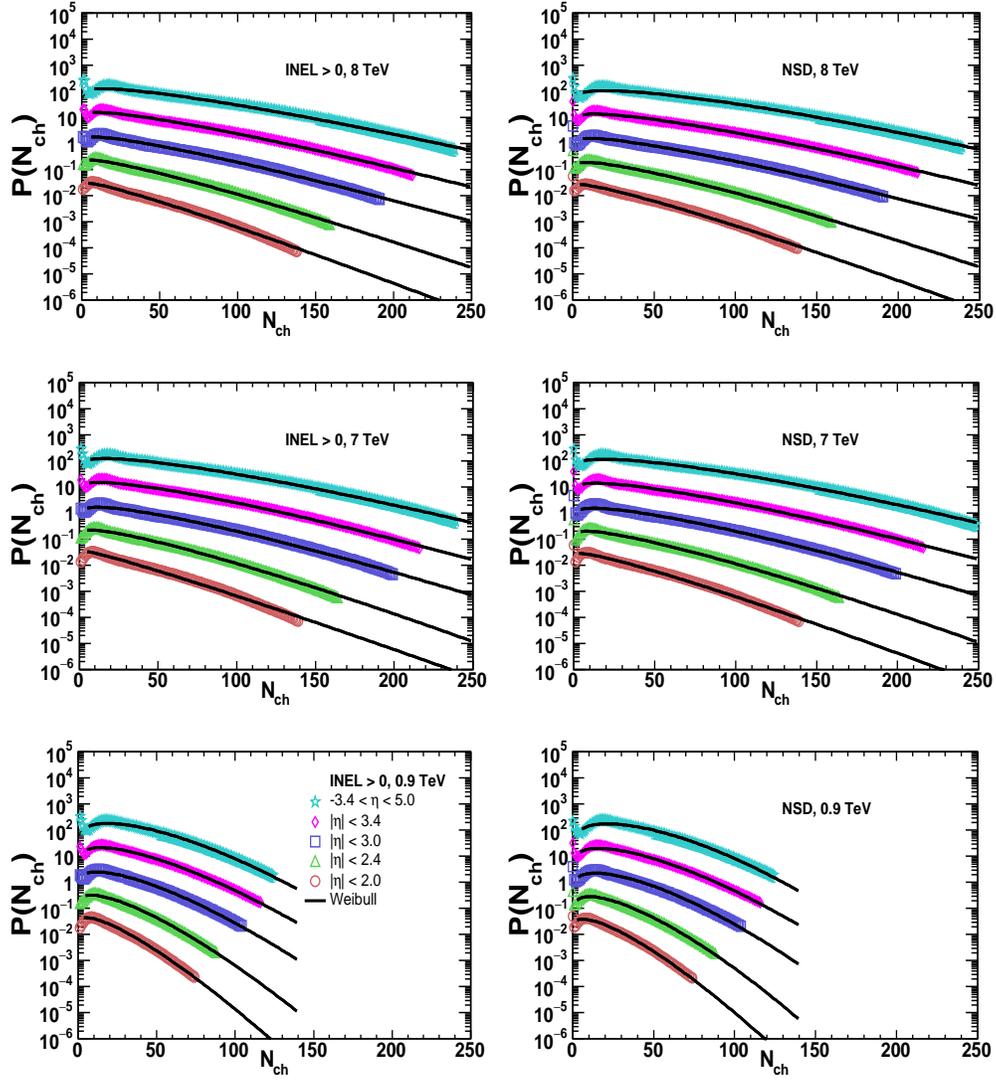} 
\caption{(Color online) The charged particle multiplicity distribution 
  measured by ALICE experiment in different $\eta$ intervals for p$-$p 
  collisions at $\sqrt{s}$ = 0.9 TeV, 7 TeV, and 8 TeV\cite{alicefor}.The solid lines show the Weibull fit to the measured data.} 
\label{fig1}
%\end{center}
\end{figure*}

\begin{figure*}
\includegraphics[width=5.2in,height=3.8in]{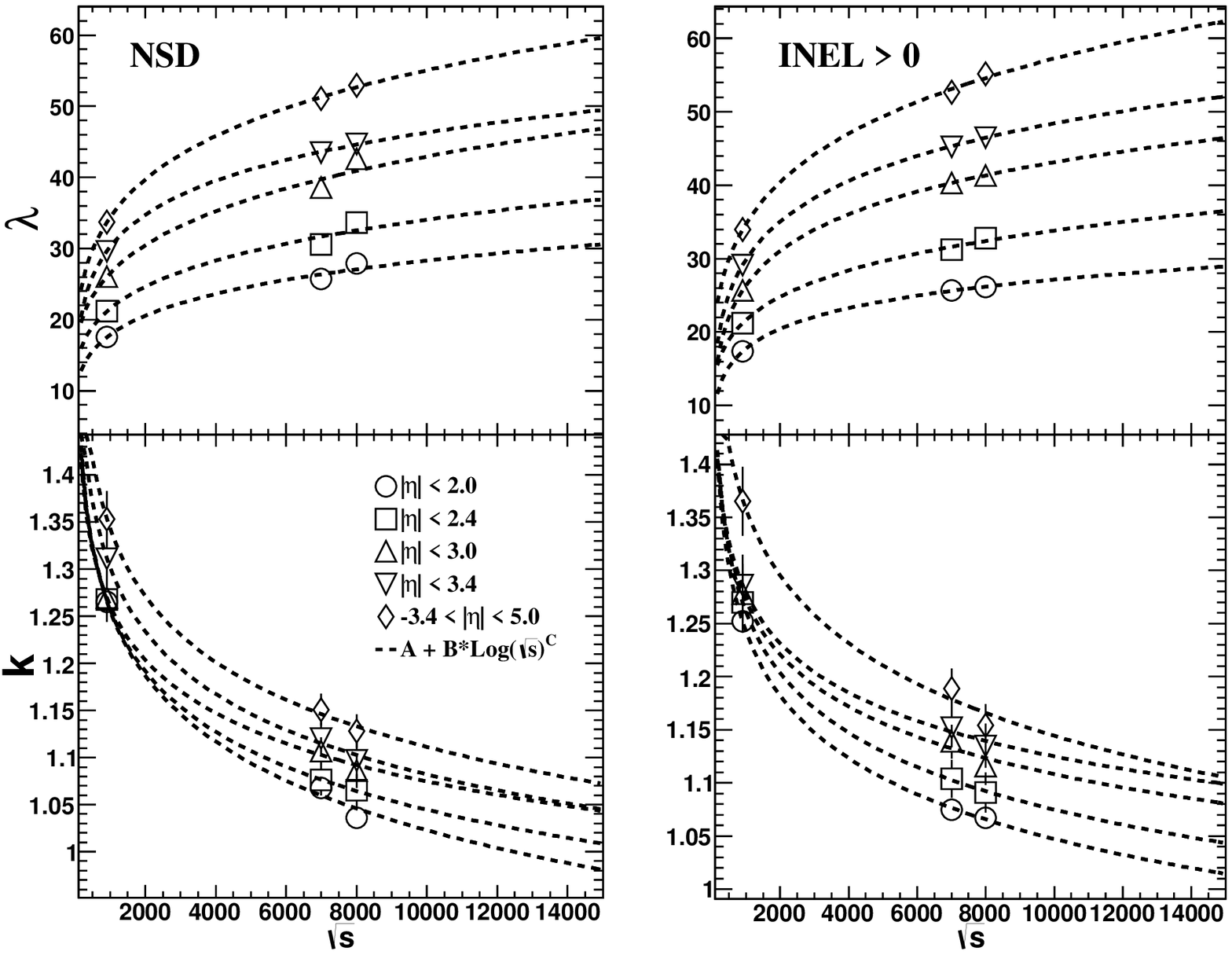} 
\caption{The variation of Weibull parameters $k$ and $\lambda$ with
  center of mass energy in p$-$p collisions. The variation 
is parameterized by a power law of the form A + B * Log $(\sqrt{s})^C$, shown by the dashed curve.}
\label{fig2}
\end{figure*}

The present work is based on the analysis of the ALICE experiment
which has measured the charged particle multiplicity distribution over
a wide $\eta$ interval(-3.4$<\eta<$5.0) in p$-$p collisions at
$\sqrt{s}$ = 0.9, 7, and 8 TeV at the LHC\cite{alicefor}. The measurement
was done for three classes of events. The first class consisted of all
inelastic collisions($INEL$) while the second class known as $INEL>0$ 
selects the same events as the first one with an additional
requirement of at least one track in the region $|\eta| <$ 1.0. The
third class  is the non-single diffractive ($NSD$) class where the
number of single-diffractive events were greatly reduced by
requiring the detection of charged particles in both the odometers.
The extension in $\eta$ coverage also enabled the measurement to reach
higher multiplicity and a different phase space when compared to
previous results published by ALICE\cite{alicemid}. 
The present analysis only considers two different event classes namely
the $NSD$ and $INEL>0$.The multiplicity distributions in two different
event classes are fitted with Weibull distribution for five different $\eta$ intervals
i.e $|\eta|<$ 2.0, $|\eta|<$ 2.4, $|\eta|<$ 3.0, $|\eta|<$ 3.4 and
-3.4$<\eta<$5.0. Figure \ref{fig1} shows the
multiplicity distributions fitted with Weibull function for 0.9 TeV, 7
TeV, and 8 TeV, respectively. The first few bins are excluded from the fits as
there is an increase in the number of events due to diffractive
events. The Weibull distribution seems to describe the data quite
well. Table \ref{table1} gives the details of the fitting parameters
and the quality of fit. This is in agreement with the  previous results
where the fit described the data well for central rapidity\cite{weib1}.
The variation of $\lambda$ and $k$ as a function of beam energy is
shown for the studied $\eta$ intervals in Figure \ref{fig2}. One can observe that the
values of $\lambda$ increase with beam energy as well as with 
the width of $\eta$ interval. The $k$ values slightly decrease with
an increase of collision energy while it shows an increasing trend
with an increase in width of $\eta$ interval. The trend of both $\lambda$
and $k$ with beam energy over all $\eta$ intervals is similar to
previous observations\cite{weib1,weib2,weibmom}.  
\begin{table*}
\centering
\begin{tabular}{c c c c c c}
\hline
$\sqrt{s}(TeV)$  &Event class & $\eta$ Range   & $k$           & $\lambda$       &  $\chi^{2}/ndf$ \\

 %TeV &           &           &      &\\
\hline 
\hline
0.9  &  NSD &  $|\eta| <$ 2.0   &  1.265 $\pm$ 0.007   &  17.54 $\pm$ 0.091 & 1.56\\
       &  &  $|\eta| <$ 2.4   &  1.2676 $\pm$ 0.01   &  21.16 $\pm$ 0.368 & 1.01\\
       &  &  $|\eta| <$ 3.0   &  1.2695 $\pm$ 0.013   &  26.028 $\pm$ 0.42 & 0.703\\
       &  &  $|\eta| <$ 3.4   &  1.31 $\pm$ 0.012   &  29.59 $\pm$ 0.56 & 0.82\\
       &  &  -3.4$ < |\eta| <$ 5.0   &  1.35 $\pm$ 0.012   &  33.68 $\pm$ 0.61 & 0.91\\
       &  INEL$>$0 &  $|\eta| <$ 2.0   &  1.25 $\pm$ 0.007   &  17.36 $\pm$ 0.11 & 1.8\\
       &  &  $|\eta| <$ 2.4   &  1.269 $\pm$ 0.01   &  21.23 $\pm$ 0.481 & 1.77\\
       &  &  $|\eta| <$ 3.0   &  1.248 $\pm$ 0.013   &  25.65 $\pm$ 0.53 & 0.43\\
       &  &  $|\eta| <$ 3.4   &  1.286 $\pm$ 0.012   &  29.09 $\pm$ 0.63 & 0.62\\
       &  &  -3.4$ < \eta <$ 5.0   &  1.365 $\pm$ 0.012   &  33.95 $\pm$ 0.69 & 1.06\\
\hline 
\hline
7    & NSD &  $|\eta| <$ 2.0   &  1.068 $\pm$ 0.007   &  25.72 $\pm$ 0.224 & 1.37\\
      & &  $|\eta| <$ 2.4   &  1.07 $\pm$ 0.01   &  30.58 $\pm$ 0.37 & 0.7\\
      & &  $|\eta| <$ 3.0   &  1.107 $\pm$ 0.013   &  38.58 $\pm$ 0.57 & 0.45\\
      & &  $|\eta| <$ 3.4   &  1.12 $\pm$ 0.013   &  43.52 $\pm$ 0.84 & 0.62\\
      & &  -3.4$ < \eta <$ 5.0   &  1.151 $\pm$ 0.014   & 51.05 $\pm$0.9 &0.419\\
      &  INEL$>$0 &  $|\eta| <$ 2.0   &  1.075 $\pm$ 0.007   &  25.65 $\pm$ 0.21 & 1.45\\
      &  &  $|\eta| <$ 2.4   &  1.104 $\pm$ 0.009   &  31.24 $\pm$ 0.33 & 0.89\\
      &  &  $|\eta| <$ 3.0   &  1.15 $\pm$ 0.01   &  40.34 $\pm$ 0.67 & 0.64\\
      &  &  $|\eta| <$ 3.4   &  1.17 $\pm$ 0.011   &  45.27 $\pm$ 0.9 & 0.57\\
      &  &  -3.4$ < \eta <$ 5.0   &  1.18 $\pm$ 0.013   &  52.88 $\pm$ 0.97 & 0.52\\
\hline
\hline
8   & NSD &  $|\eta| <$ 2.0   &  1.03 $\pm$ 0.007   &  27.93 $\pm$ 0.23 & 1.03\\
      & &  $|\eta| <$ 2.4   &  1.06 $\pm$ 0.01   &  33.61 $\pm$ 0.39 & 0.72\\
      & &  $|\eta| <$ 3.0   &  1.08 $\pm$ 0.01   &  42.62 $\pm$ 0.44 & 0.67\\
      & &  $|\eta| <$ 3.4   &  1.09 $\pm$ 0.01   &  44.7 $\pm$ 0.57 & 0.35\\
      & &  -3.4$ < \eta <$ 5.0   &  1.12 $\pm$ 0.013   &  52.93 $\pm$0.68 &0.46\\
      &  INEL$>$0 &  $|\eta| <$ 2.0   &  1.06 $\pm$ 0.007   &  26.13 $\pm$ 0.22 & 1.1\\
      &  &  $|\eta| <$ 2.4   &  1.09 $\pm$ 0.011   & 32.82 $\pm$ 0.7 & 0.62\\
      &  &  $|\eta| <$ 3.0   &  1.116 $\pm$ 0.013   &  41.33 $\pm$ 0.78 & 0.54\\
      &  &  $|\eta| <$ 3.4   &  1.13 $\pm$ 0.02   &  46.52 $\pm$ 1.04 & 0.48\\
      &  &  -3.4$ < \eta <$ 5.0   &  1.15 $\pm$ 0.02   &  55.15 $\pm$ 1.1 & 0.42\\
\hline
\end{tabular}
\caption{ \label{table1} The values of  $k$, $\lambda$ and
  $\chi^{2}/ndf$ obtained from the fits of multiplicity distributions using Weibull
  function in p$-$p collisions for different $\eta$ intervals at
  different energies as measured by ALICE experiment at LHC \cite{alicefor}.} 
\end{table*}

\begin{figure*}
\includegraphics[width=5.1in,height=3.2in]{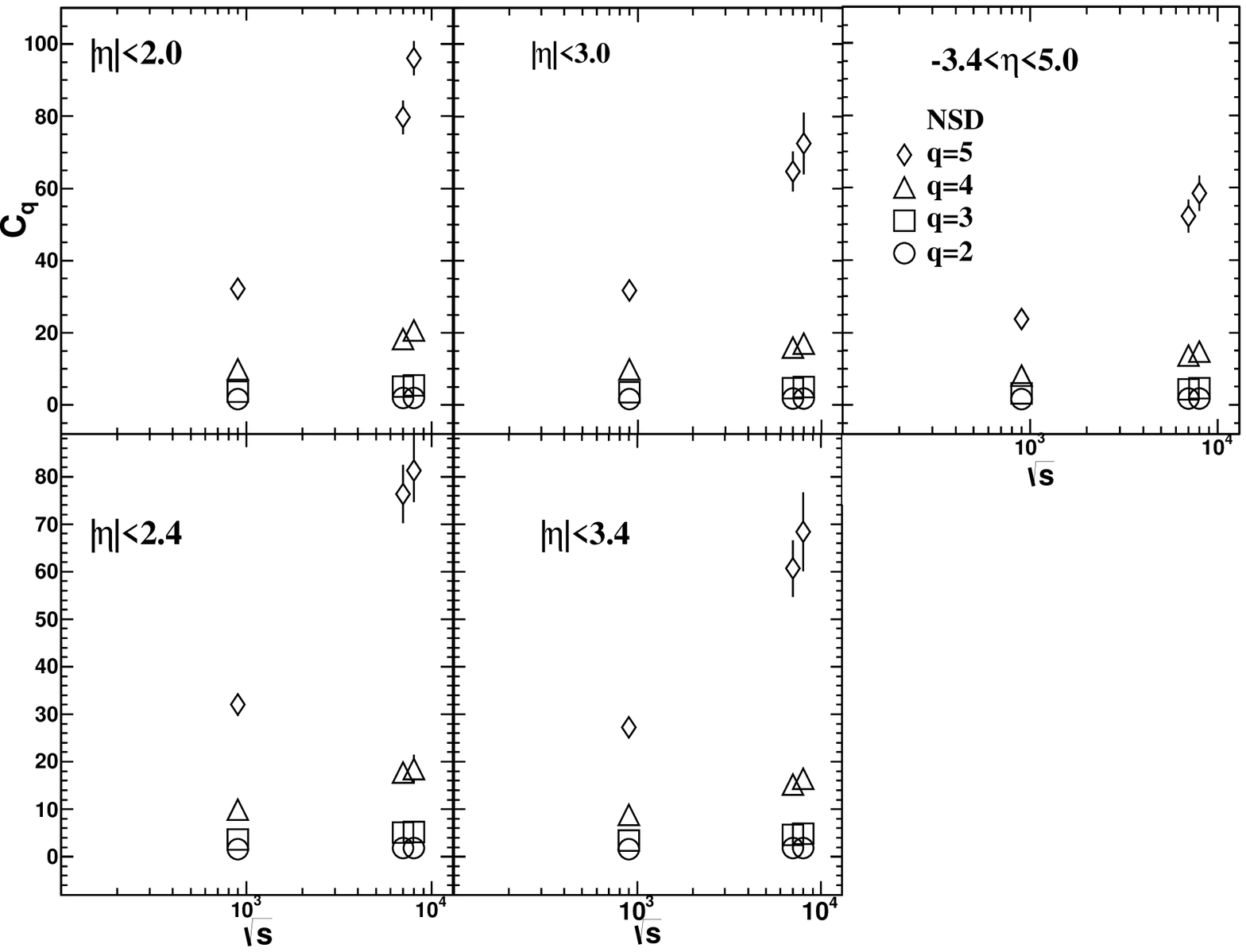} 
\caption{The variation of the normalized moments for $NSD$ event class
 with collision energy as obtained from Weibull calculations for
 different $\eta$ intervals}
\label{fig3}
\end{figure*}

\begin{figure*}
\includegraphics[width=5.1in,height=3.2in]{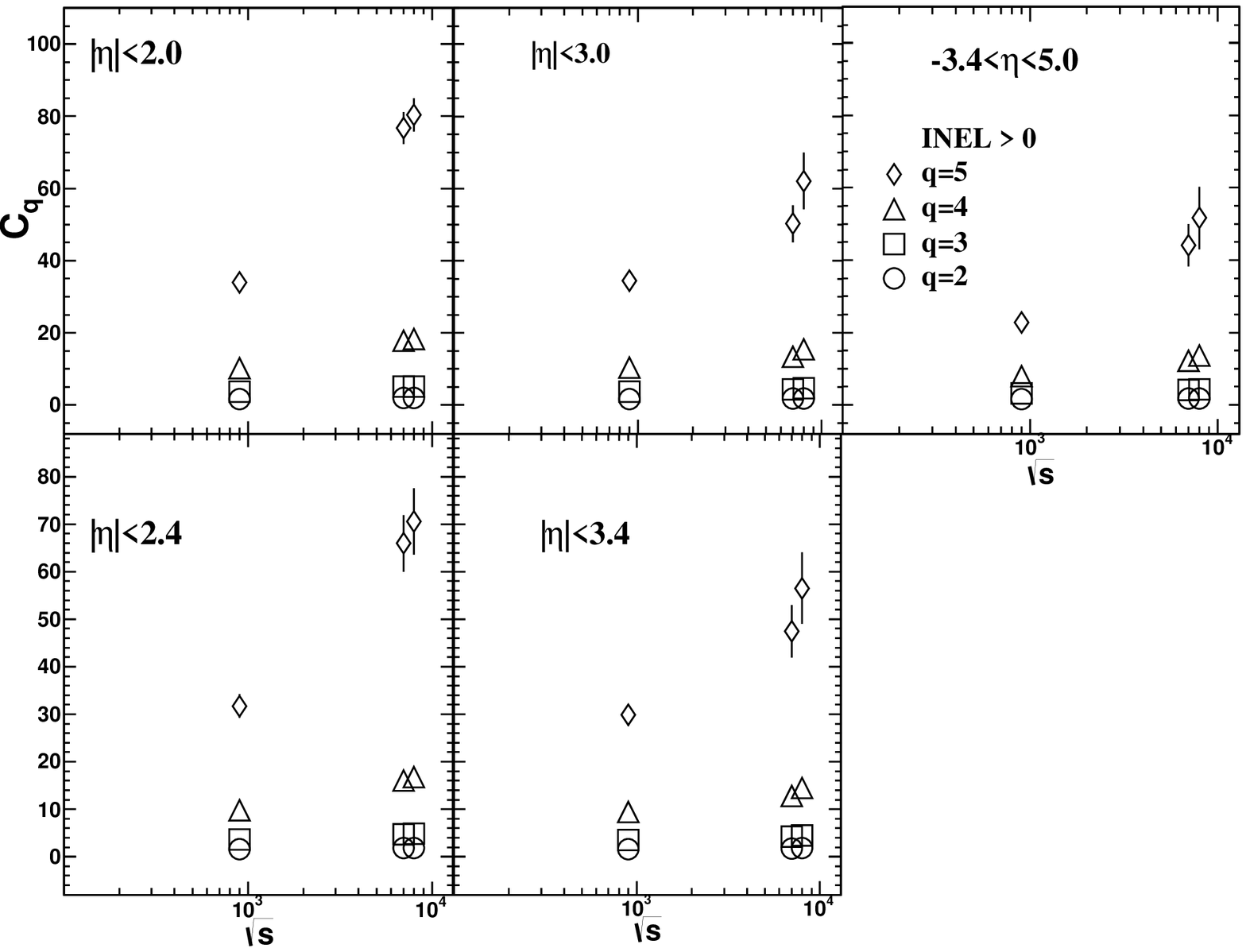} 

\caption{The variation of the normalized moments for $INEL > 0$ event class
 with collision energy as obtained from Weibull calculations for
 different $\eta$ intervals.}
\label{fig4}
\end{figure*}

\begin{figure*}
\includegraphics[width=5.1in,height=3.2in]{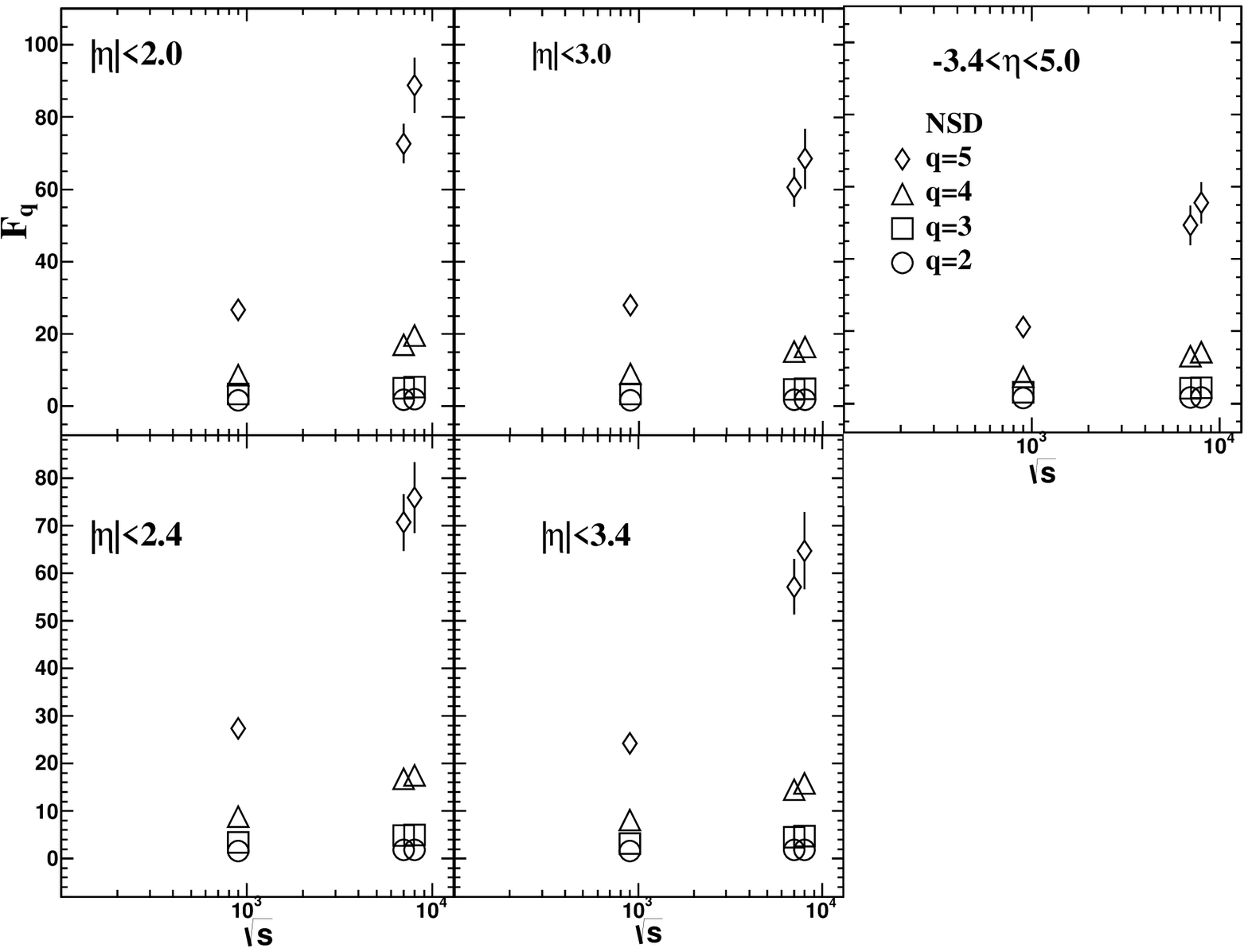} 
\caption{The variation of the normalized factorial moments for $NSD$
  event class with collision energy as obtained from Weibull calculations for
 different $\eta$ intervals.} 
\label{fig5}
\end{figure*}

\begin{figure*}
\includegraphics[width=5.1in,height=3.2in]{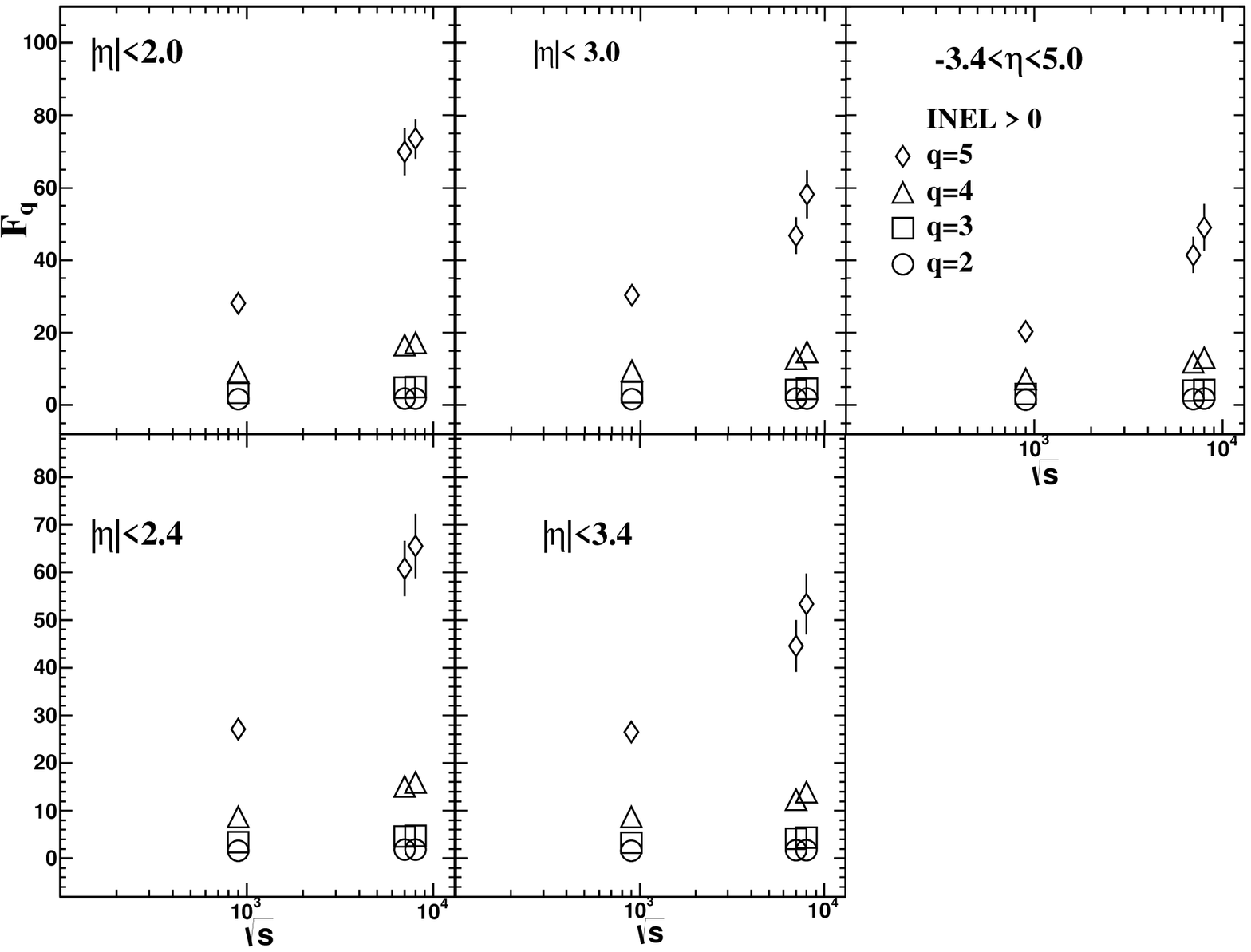} 
\caption{The variation of the normalized factorial moments for $INEL >
  0$ event class with collision energy as obtained from Weibull calculations for
 different $\eta$ intervals.}
\label{fig6}
\end{figure*}

The normalized raw moments and the factorial moments (up to fifth
order) have also been calculated using the extracted parameters using the
equation 5. Fig \ref{fig3} shows the variation of normalized moments
($C_{n}$) for NSD and Fig \ref{fig4} shows the same
for $INEL>0$  eventclass. Fig\ref{fig5} and \ref{fig6} shows the
variation of factorial moments ($F_{n}$) with beam energy for
different $\eta$ intervals for NSD  and $INEL>0$ eventclass,
respectively. It can be seen for all pseudorapidity intervals studied,
$C_{2}$(and $F_{2}$) remains constant with energy while $C_{3}$(and
$F_{3}$) shows a small increase with increasing energy. However,
$C_{4}$ (and $F_{4}$) and $C_{5}$ (and $F_{5}$) show an increase with
increasing energy, which becomes stronger for larger $\eta$
intervals. This is observed in both the event classes and signals
towards the violation of KNO scaling in wider $\eta$ intervals. The 
observation is in agreement with the previous experimental results.
The values of $k$ and $\lambda$  for p$-$p collisions at 13 TeV
are extrapolated from the parametrization of the variation of the
parameters as shown in Fig \ref{fig2}. The estimated  values of normalized 
moments and factorial moments for different $\eta$ ranges are listed
in Table \ref{table2}. The predictions are important for future
experimental measurements of multiplicity distribution and its moments 
in the forward region for p$-$p collisions at $\sqrt{s} = $ 13 TeV. 

\begin{table*}
\centering
\begin{tabular}{c c c c c c c c c c c c}
\hline
Event class & $\eta$ Range   & $k$    & $\lambda$  &$C_{2}$ & $C_{3}$
  &$C_{4}$ &$C_{5}$  &$F_{2}$  &$F_{3}$  &$F_{4}$ &$F_{5}$ \\
\hline 
\hline
 NSD  & $|\eta| <$ 2.0   & 30.447   &  0.987  & 2.026 &6.201 &25.411
           &130.597  &  1.994   &6.005 &24.2202   &122.526  \\
         & $|\eta| <$ 2.4   & 36.643   &  1.021  & 1.960 &5.70 &21.958
           &105.193  &  1.932   &5.540 &21.0328   &99.297  \\
          & $|\eta| <$ 3.0   & 46.731   &  1.046  & 1.914 &5.378 &19.855   &90.6133       
         &  1.893   &5.254 &19.16   &86.376  \\
        & $|\eta| <$ 3.4   & 48.463   &  1.051  & 1.906 &5.318 &19.471   &88.0322       
          &  1.885   &5.199 &18.80   &84.0159  \\
        & -3.4$<\eta <$ 5.0   & 57.814   &  1.08  & 1.859 &5.318 &19.471   &88.0322              
       &  1.841   &4.896 &16.93   &71.827  \\
\hline
\hline
 INEL$>$0  & $|\eta| <$ 2.0   & 28.233   &  1.030  & 1. 943 &5.580 &21.163   &99.6 
                  &  1.907   &5.374 &19.99   &92.259  \\
                  & $|\eta| <$ 2.4   & 35.652   &  1.053  & 1.903 &5.295 &19.321   &87.028 
                  &  1.874   &5.133 &18.428   &81.646  \\
                  & $|\eta| <$ 3.0   & 45.223   &  1.081  & 1.857 &4.984 &17.39   &74.479       
                 &  1.834   &4.858 &16.72   &70.605  \\
                 & $|\eta| <$ 3.4   & 50.862   &  1.103  & 1.824 &4.764 &16.08   &66.307       
                 &  1.804   &4.653 &15.509   &63.095  \\
                 & -3.4 $<\eta <$ 5.0   & 60.50   &  1.108  & 1.815 &4.708 &15.75   &64.295   
                 &  1.798   &4.615 &15.27   &61.63  \\
 \hline 
\end{tabular}
\caption{ \label{table2} The values of $\lambda$, $k$, normalized
  moments and factorial moments of  various order obtained from the
  fits of multiplicity distributions using Weibull function in p$-$p
  collisions for different $\eta$ intervals at 13 TeV.} 
\end{table*}

\section{Conclusion}

The charged-particle multiplicity in forward rapidity as measured by
ALICE experiment for $\sqrt{s}$ = 0.9 TeV, 7 TeV, and 8 TeV is well
described by the two parameter Weibull distribution. The extracted
fitting parameters were used to calculate the normalized moments and
the factorial moments (up to 5th order) of the distribution for
various $\eta$ intervals. The model predictions of the higher moments
indicate towards a strong violation of KNO scaling in the forward
region in the two considered event classes. This observation is in agreement with
the previous measurement at central rapidity.

\end{document}